# Spatial-enhanced Reflective Coded Aperture Snapshot Spectral Imaging

Jiayu Di[1,*]   Yantao Jin[1,*]   Zhenming Yu[1,+]   Liming Cheng[1]
Jingyue Ma[1]   Ning Zhan[1]   Kun Xu[1]

[1]Beijing University of Posts and Telecommunications
yuzhenming@bupt.edu.cn

**Abstract**

*Coded aperture snapshot hyperspectral imaging (CASSI) system which captures 2-D spatial information and 1-D spectral information in just one or two shots has become a promising technology to capture hyperspectral image (HSI). However, previous CASSI have shortcomings such as poor spatial resolution and low light efficiency that hinder their further applications. In this paper, we propose a spatial-enhanced reflective coded aperture snapshot spectral imaging system (SE-RCASSI). The system achieves superior spatial results and high light efficiency because of its specially designed structure. Then, we propose Spatial-enhanced Network (SEnet) which takes full use of the prior information of grayscale image to boost the reconstruction quality and can meet different requirements in applications. Furthermore, we propose hybrid prior strategy (HPS) to effectively exploit the broad-scope prior of training set and narrow-scope prior of coded input, resulting in improved generalization and performance of the network. Finally, we fabricate the prototype of SE-RCASSI and conduct experiments in different environments. Both experimental results and numerical simulations show the outstanding performance of our method.*

## 1. Introduction

With the rapid development of autonomous driving and remote sensing, multi-dimensional images are highly demanded for conducting various detection tasks. Hyperspectral image (HSI) which contains 2D spatial information and 1D spectral information is a great choice for the usage of object detection, image classification, food inspection and remote sensing. Traditional spectral imaging systems have shortcomings such as large volume, low acquisition speed, poor image quality and so on which limit their further applications. By leveraging the compressive sensing theory, coded aperture snapshot hyperspectral imaging (CASSI) system has become a compromising imaging system for HSI acquisition because of its high acquisition speed.

[*]Equal Contribution,[+]Corresponding Author

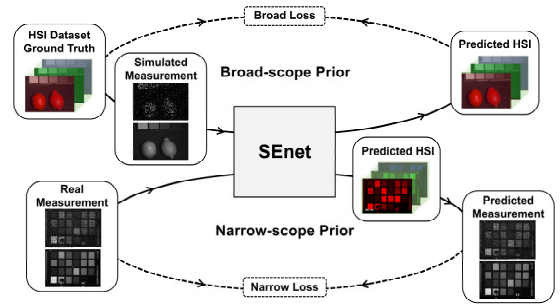

Fig.1. Overview of hybrid prior strategy.

CASSI systems are mainly divided into two categories, one is single-disperser CASSI (SD-CASSI), and the other is dual-disperser CASSI (DD-CASSI). DD-CASSI encodes the light by a fixed mask then shears the light twice, ending up in better spatial resolution than SD-CASSI. However, the DD-CASSI has a complex optical structure, which leads to hardship in building the system. Reflective coded aperture snapshot spectral imaging system (RCASSI) could achieve as great spatial resolution as DD-CASSI with fewer optical devices, while only half of the light is utilized, leading to the poor light efficiency.

Besides the hardware of the hyperspectral imaging system, the quality of the hyperspectral images also depends on the reconstruction algorithms. Model-based methods and deep learning-based methods are proposed to reconstruct HSI. Model-based algorithms which use hand-crafted prior to reconstruct HSI always have poor spatial result and have limitations in practical applications. Algorithms based on deep learning can reach accurate HSI reconstruction by learning the prior information in datasets. However, those deep learning-based algorithms often have limited performance when processing unknown data, which hinders the deployment of algorithms in further applications. Furthermore, the prior information of coded input is always ignored in previous deep learning-based algorithms. Generalization of the algorithms could be greatly increased by utilizing the prior information of coded input.

In this paper, we propose the spatial-enhanced reflective coded aperture snapshot spectral imaging (SE-RCASSI) system. SE-RCASSI is an improved version of RCASSI. Compared to RCASSI, SE-RCASSI utilizes the light that remains unused in RCASSI, contributing to a great improvement of light efficiency. The SE-RCASSI also has the advantages of high spatial resolution for additional path acquiring additional spatial information of the scene. The structure of SE-CASSI in shown in Fig.2 (a). Furthermore, we propose a HSI reconstruction method called Spatial Enhancement Network (SEnet) to fit the system. Its specially designed modules enable the network to efficiently fuse coded measurement with grayscale measurement, resulting in better reconstruction quality. Particularly, the pre-reconstruction module of SEnet can be substituted with different deep learning-based reconstruction algorithms. By deploying advanced pre-reconstruction algorithms, the SEnet can achieve more accurate fusion results. Various requirements in different application can also be met by deploying corresponding algorithms. Finally, we propose a hybrid prior strategy (HPS) which combines broad-scope prior of training set with narrow-scope prior of coded measurement to achieve more precise reconstruction and boost of generalization, as shown in Fig. 1. The main contributions are summarized as follows:

(1) We propose the SE-RCASSI system for HSI capture, which performs well in acquisition speed, spatial resolution, and light efficiency.

(2) Based on the characteristics of the measurement, we propose SEnet together with HPS to improve the reconstruction performance and increase the generalization of the model.

(3) We fabricate a miniaturized SE-RCASSI system prototype and conduct experiments in various environments.

## 2. Related work

### 2.1. Coded aperture snapshot compressed imaging.

Snapshot hyperspectral imaging which captures 2D spatial image and 1D spectral information in just one shot is a distinguished tool in fields related to object tracking, remote sensing, environment monitoring and so on. CASSI is a compromising snapshot compressed imaging system that has attracted widespread attention due to its simple process of system setup and calibration. Conventional CASSI encodes the light by using a binary-coded aperture to block or unblock the light, thus having shortcomings such as poor light efficiency and low spatial resolution. To improve the structure and reconstructed quality of CASSI, different methods have been proposed by previous studies. Yu et al. proposed the RCASSI which reduces the device and volume by half while still maintaining the quality of acquisition results. However, half of the incident light in RCASSI is ignored, resulting in poor light efficiency. Wang et al. proposed a dual-camera compressed hyperspectral imaging system (DCCHI) which takes use of an extra panchromatic camera to acquire more spatial information of the scene. However, DCCHI employ structure of SD-CASSI to encode the light, which limits the spatial resolution of the system. Jonathan et al. utilized the digital micro-mirror device (DMD) and a dispersive prism to disperse-code-disperse the light, achieving better reconstruction results than DD-CASSI. However, multiple shots are required in the system, making the system lose the advantage of snapshot. Furthermore, various modifications have been made in traditional CASSI to improve the acquisition accuracy. In conclusion, the previous CASSI-based systems still have limitations in spatial resolution, light efficiency, or acquisition speed. The practicality of CASSI systems in various applications is enormously limited by those weaknesses.

### 2.2. Reconstruction algorithm.

HSI reconstruction for SCI system is mathematically an ill-posed problem that needs prior information. Different model-based iterative optimization methods utilize hand-crafted prior to solve this problem. For example, the two-step iterative thresholding (TwIST) algorithm and generalized alternating projection (GAP) algorithm used total variation (TV) prior as regularization term to solve the problem. Many algorithms used sparsity prior to reconstruct HSI which can preserve image boundary and smooth region. However, model-based methods often have long run time and face lack of spatial details.

Recently, deep learning has become a powerful tool in the field of HSI reconstruction. For example, Convolutional Neural Network (CNN) can efficiently learn image priors and features during the training process. Instead of designing complex hand-craft priors, deep learning-based methods can reconstruct very accurate HSI in short time. Yu et al. proposed the Unet-3D network which employ a 3D convolution kernel to exploit the correlation of adjacent spectral channels. Meng et al. used three self-attention modules in a CNN-based framework to acquire the dependencies in spatial or spectral dimensions. Yuan firstly deploy transformer in the task of HSI reconstruction, exploiting the long-range dependencies in spectral dimension. However, deep learning-based methods still have weakness in processing data which different from training set.

Image fusion is a technique that combines multiple images from different sources, which can be used in HSI reconstruction to improve the image quality. Wang et al. firstly fused compressed image from SD-CASSI with panchromatic image and used model-based method to reconstruct HSI, achieving high reconstruction quality. Xie et al. leverage the self-supervised framework and transformer to achieve the fuse and HSI reconstruction. However, previous methods based on image fusion to improve the HSI reconstruction quality still suffer from complex training process and long run time. End to end image fusion network has not been propose in the field of HSI reconstruction till now.

image of scenes. Since grayscale measurements are not encoded, they have more spatial details than compressed measurements.

### 3.2. Observation Model

Let $\mathbf{X}(m,n,\lambda)$ indicates the intensity of incident light where $1 \leq m \leq M$ and $1 \leq n \leq N$ are the spatial coordinates and $1 \leq \lambda \leq \Lambda$ represents the spectral coordinate. The

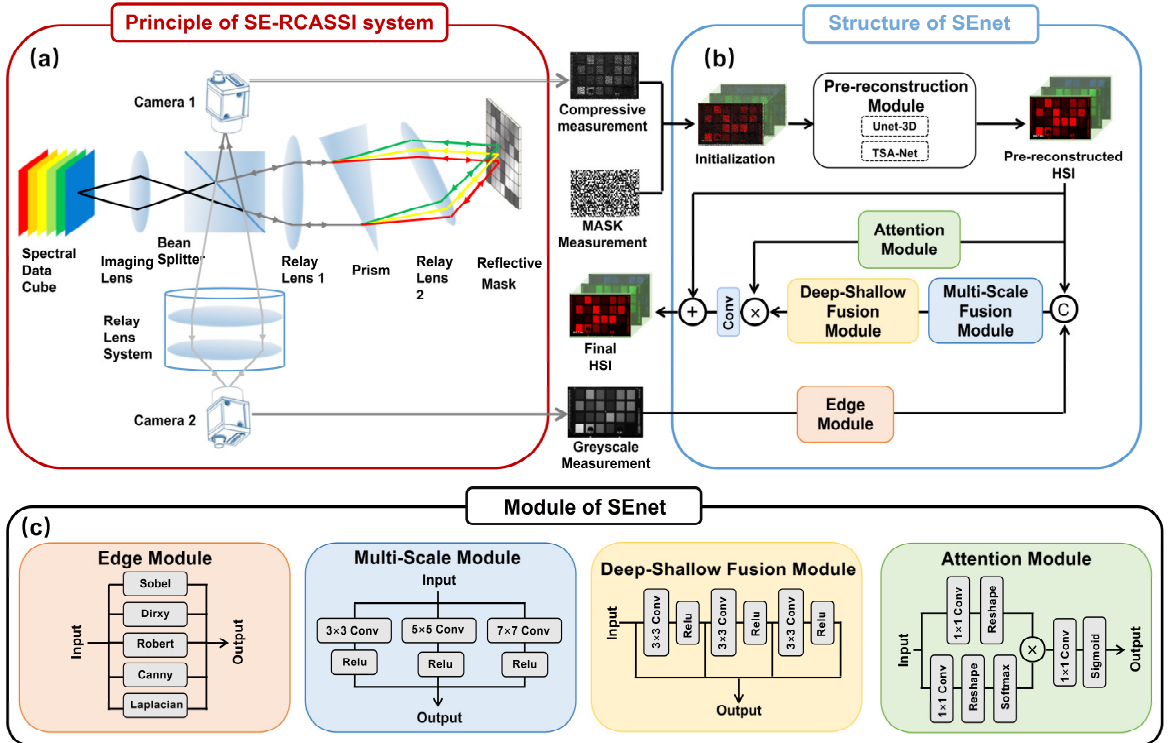

Fig.2. The overall architecture of SE-RCASSI and SEnet. (a) Principle of SE-RCASSI system. (b) Structure of SEnet. (c) Architecture of the modules.

## 3. Methods

### 3.1. System setup

Fig. 2 depicts the SE-RCASSI system, with panel (a) detailing the principle of the optical system and panel (b) showing the structure of SEnet. The architecture of different modules in SEnet is shown in Fig. 2(c). The optical system consists of an imaging lens, two relay lenses, a relay lens system, a beam splitter, a prism, two monochrome cameras, and a reflective random mask. The light is focused by the objective lens and then reaches the beam splitter. The beam splitter allows half of the light to pass through and reflects the other half, thereby forming two distinct light paths. The SE-RCASSI system employs two identical cameras to capture information of both paths. The transmitted light is utilized for the system's main path, which undergoes compressive sampling via the random reflective mask. The side path directly captures grayscale

$(m,n)\text{-}th$ pixel of the compressed measurement of the main path in this system can be represented as:

$$y_{\text{main}}(m,n) = \sum_{\lambda=1}^{\Lambda} \varphi(m-\psi(\lambda),n)x(m,n,\lambda)\omega(\lambda) \quad (1)$$

where $\varphi(m,n)$ denotes the transmission function of the coded aperture, $\psi(\lambda)$ represents the wavelength-dependent dispersion function for the prism, $x$ is the spectral distribution of the $(m,n,\lambda)$–th pixel of the HSI, $\omega(\lambda)$ represents the response function of the detector.

The observation model of main path can be rewritten in matrix form as:

$$\mathbf{Y}_{\text{main}} = \mathbf{H}_{\text{main}}\mathbf{X} \quad (2)$$

where $\mathbf{H}_{main}$ denotes the projection matrix of main path and jointly determined by $\varphi(m,n)$, $\psi(\lambda)$ and $\omega(\lambda)$, $\mathbf{Y}_{main}$ denotes the vectorized representation of the compressed measurement $y_{main}(m,n)$, $\mathbf{X}$ is the HSI.

The $(m,n)$-$th$ pixel of the grayscale measurement can be represented as:

$$y_{\text{gray}}(m,n) = \sum_{\lambda=1}^{\Lambda} x(m,n,\lambda)\omega(\lambda) \qquad (3)$$

Similar with Equation (2), Equation (3) can also be rewritten in matrix form as

$$\mathbf{Y}_{gray} = \mathbf{H}_{gray}\mathbf{X} \qquad (4)$$

where $\mathbf{H}_{gray}$ denotes the projection matrix of grayscale camera and determined by $\omega(\lambda)$, $\mathbf{Y}_{gray}$ is the vectorized representation of the grayscale image.

Consequently, the SE-RCASSI sensing process can be generally expressed as:

$$\mathbf{Y} = \mathbf{H}\mathbf{X} + e \qquad (5)$$

Where:

$$\mathbf{H} = \begin{bmatrix} \mathbf{H}_{main} \\ \mathbf{H}_{gray} \end{bmatrix}, \mathbf{H} = \begin{bmatrix} \mathbf{H}_{main} \\ \mathbf{H}_{gray} \end{bmatrix}, e = \begin{bmatrix} e_{main} \\ e_{gray} \end{bmatrix} \qquad (6)$$

$e_{main}$ and $e_{gray}$ represent the additive noise of the two detectors, which are generally modeled as white gaussian noise.

Given the system sensing matrix $\mathbf{H}$, our goal is recovering $\mathbf{X}$ from the incomplete measurements $\mathbf{Y}$. Exploring the best solution, we propose SEnet, as shown in Fig. 2.

### 3.3. Network structure

The measurements of SE-RCASSI consist of three parts: compressed measurement, mask measurement, and grayscale measurement. The initialization of HSI is obtained by multiplying the compressed measurement with the mask measurement. Then, the initialized data is sent into a pre-reconstruction module, which is a deep neural network that can recover a hyperspectral image from the compressed measurement. The preliminarily reconstructed HSI suffers from low spatial resolution and blurring artifacts, owing to the balance between spectral and spatial information in compressed measurements.

To improve the spatial quality by fusing the grayscale measurement, we propose several modules in SEnet. The SEnet is designed to extract spatial features from the grayscale measurement and generate a final output that preserves both spectral and spatial details. In addition to the HSI pre-reconstruction module, our SEnet comprises other four modules: multi-scale module, attention module, deep-shallow fusion module, edge module. These modules work together to enhance the spatial quality of hyperspectral images while maintaining spectral fidelity. In the following we detail each part of the SEnet, as shown in Fig. 2(c).

**Edge Module.** In simulations and experiments, we observe that the high-frequency details such as edges, contours, and textures in the pre-reconstructed HSI are significantly inferior to those in the grayscale measurement. This observation highlights the potential to exploit the high-frequency details in the grayscale measurement, rather than depending solely on the original image for training. The output of the edge module is generated by combining the grayscale measurement and high-frequency features. We use five common high-frequency operators to extract high-frequency features. This process can be formulated as:

$$\begin{aligned}\mathbf{O} = [&\alpha_{dir-xy}\mathbf{Y}_{gray}, \alpha_{robert}\mathbf{Y}_{gray}, \alpha_{canny}\mathbf{Y}_{gray},\\ &\alpha_{sober}\mathbf{Y}_{gray}, \alpha_{laplacian}\mathbf{Y}_{gray}, \mathbf{Y}_{gray}]\end{aligned} \qquad (8)$$

where $\alpha_{dir\text{-}xy}$, $\alpha_{robert}$, $\alpha_{canny}$, $\alpha_{sober}$, $\alpha_{laplacian}$ represent first-order difference operator, Robert operator, Canny operator, Sobel operator, and Laplacian operator, respectively.

**Multi-Scale Module.** The output of the edge module and the pre-reconstructed HSI are concatenated along the spectral dimension as the input of the multi-scale module. The multi-scale convolutional module has been extensively applied in the fields of deep learning and computer vision, particularly in tasks related to image processing and segmentation. By processing images at multiple scales, the module is able to capture features ranging from fine details to large-scale structures. The module employs three convolutional kernels of distinct sizes to extract features within varying receptive fields. This process can be formulated as:

$$\begin{cases} \mathbf{O}_3 = ReLu\left(\mathbf{C}_3 * \left[\mathbf{O}_{gray}, \mathbf{O}_{reco}\right]\right) \\ \mathbf{O}_5 = ReLu\left(\mathbf{C}_5 * \left[\mathbf{O}_{gray}, \mathbf{O}_{reco}\right]\right) \\ \mathbf{O}_7 = ReLu\left(\mathbf{C}_7 * \left[\mathbf{O}_{gray}, \mathbf{O}_{reco}\right]\right) \\ \mathbf{O}_m = [\mathbf{O}_3, \mathbf{O}_5, \mathbf{O}_7] \end{cases} \qquad (9)$$

where $\mathbf{C}_i$, $Relu$, and $\mathbf{O}_{reco}$ represent the convolutional layer, the ReLU activation function and CASSI reconstructed HSI, respectively. $\mathbf{O}_i$ is the output of the response convolutional layer, the subscript $i$ ($i$ =3, 5, 7) means the size of the convolutional kernel, $\mathbf{O}_m$ is the output of this multi-scale module.

**Deep-Shallow Fusion Module.** In CNN, shallow features usually refer to edges and textures. As the network goes deeper, deep features begin to represent higher-level semantic content of images, containing more complex patterns and structures. In order to utilize the advantages of both shallow and deep layers effectively, we design the deep-shallow fusion module. This module combines shallow and deep features to obtain a more comprehensive reconstruction of the image. Therefore, this process can be formulated as:

$$\begin{cases} \mathbf{O}_{d1} = ReLu(\mathbf{C}_3 * \mathbf{O}_m) \\ \mathbf{O}_{d2} = ReLu(\mathbf{C}_3 * \mathbf{O}_{d1}) \\ \mathbf{O}_{d3} = ReLu(\mathbf{C}_3 * \mathbf{O}_{d2}) \\ \mathbf{O}_d = [\mathbf{O}_m, \mathbf{O}_{d1}, \mathbf{O}_{d2}, \mathbf{O}_{d3}] \end{cases} \quad (10)$$

where $\mathbf{O}_{di}$ and $\mathbf{O}_d$ represent the output of the $i-th$ convolution and the deep-shallow fusion module, respectively.

**Attention Module.** Given the inherent spectral information in $\mathbf{O}_{reco}$, we introduce the spectral attention mechanism to keep the spectral information from being impaired. We adopt scale reshaping and activation operations on $\mathbf{O}_{reco}$ to characterize the relationship among channels. The input is processed as:

$$\begin{cases} \mathbf{O}_{query} = \mathbf{C}_1 * \mathbf{O}_{reco} \\ \mathbf{O}_{key} = \mathbf{C}_1 * \mathbf{O}_{reco} \\ \mathbf{O}_{value} = \mathbf{F}_r(\mathbf{O}_{key}) \otimes Softmax(\mathbf{F}_r(\mathbf{O}_{query})) \\ s = \mathbf{C}_1 Sigmoid(\mathbf{O}_{value}) \end{cases} \quad (11)$$

where $\mathbf{C}_1$ is the 1 × 1 convolution kernel and $\mathbf{F}_r$ is reshape function. $\otimes$ means the matrix multiplication operations. *Softmax* and *Sigmoid* are the activation function. *s* is a vector representing the output of the attention module.

With the deployment of this module, the final output is rescaled and expressed as follows:

$$\hat{\mathbf{X}} = \mathbf{C}_3(s \odot \mathbf{O}_d) + \mathbf{O}_{reco} = f(\mathbf{Y}, \Theta) \quad (12)$$

Where $f$ is the mapping function of SEnet and $\Theta$ is the parameters of SEnet.

### 3.4. Hybrid Prior Strategy

To recover the structure and details of the image more accurately, we should make full use of these prior information. Therefore, we propose the hybrid prior strategy (HPS).

**Broad-scope Prior.** Broad-scope prior refers to prior knowledge learned from a wide range of data. It provides the model with a more comprehensive background and context, allowing it to better understand and handle various data scenarios. It is obtained through supervised learning on a wide range of data, which establishes a stable and universally applicable benchmark for image reconstruction.

Let $\mathbf{X}_{broad}$ indicate the ground truth of broad-scope set, the prediction result of the network can be represented as:

$$\hat{\mathbf{X}}_{broad} = f(\mathbf{Y}_{broad}, \Theta_{broad}) = f(\mathbf{H}\mathbf{X}_{broad}, \Theta_{broad}) \quad (13)$$

the model is optimized by minimizing the loss:

$$\mathcal{L}_{broad}(\Theta_{broad}) = \text{MSE}(\mathbf{X}_{broad}, \hat{\mathbf{X}}_{broad}) \quad (14)$$

where MSE represent mean squared error.

**Narrow-scope Prior.** Narrow-scope prior refers to prior knowledge sourced from specific or limited datasets. It is obtained by inputting measurements into a pre-trained network for self-supervised training. This approach emphasizes the distinctive characteristics of individual scenes or datasets, ensuring precise modeling of subtle variations and specific attributes in hyperspectral images.

Given the relationship between the capture image and the corresponding HSI in Eq. (2), the predicted HSI in Eq. (12) after the linear mapping $\mathbf{H}$ should be consistent with the input image. We add the spatial-spectral constraint in Equation (14) for reconstruction and the internal learning loss can be expressed as:

$$\mathcal{L}_{narrow}(\Theta_{narrow}) = \text{MSE}(\mathbf{H}\hat{\mathbf{X}}_{narrow}, \mathbf{Y}_{narrow}) \quad (15)$$

We optimize the model by minimizing the $\mathcal{L}_{narrow}$ and $\mathcal{L}_{broad}$. Narrow-scope priors and broad-scope priors are utilized in the minimization process.

## 4. Assessment

In this section, our proposed SE-RCASSI system which includes hardware optical setup and reconstruction algorithm is demonstrated. Experiments and simulations have been conducted to show the accuracy of our system.

### 4.1. Datasets and Metrics

We choose two hyperspectral image datasets, CAVE and KAIST to train and verify our model. CAVE datasets, which consists of 32 hyperspectral images at the size of 512 ×512, has been randomly rotated and spliced to augment the dataset. We use the extended CAVE dataset which is posed of 200 hyperspectral scenes at the spectral resolution of 10nm ranging from 400nm to 700nm and spatial resolution of 1024×1024 to train our network. We choose 10 scenes from KAIST dataset as test set to verify the accuracy of our algorithm, then compare our algorithm with other SOTA algorithms on test set.

| | RCASSI | | | | | | | | SE-RCASSI | | | | | | | | | | | | | | | |
|---|---|---|---|---|---|---|---|---|---|---|---|---|---|---|---|---|---|---|---|---|---|---|---|---|
| Scene | TwIST | | ADMM | | Unet-3D | | TSA-Net | | TwIST-DC | | ADMM-DC | | Unet-3D-DC | | TSA-Net-DC | | SEnet-U | | SEnet-T | | SEnet-UH | | SEnet-TH | |
| | PSNR | SSIM | PSNR | SSIM | PSNR | SSIM | PSNR | SSIM | PSNR | SSIM | PSNR | SSIM | PSNR | SSIM | PSNR | SSIM | PSNR | SSIM | PSNR | SSIM | PSNR | SSIM | PSNR | SSIM |
| 1 | 23.82 | 0.743 | 27.40 | 0.808 | 32.57 | 0.937 | 33.43 | 0.950 | 25.78 | 0.817 | 26.82 | 0.808 | 34.80 | 0.961 | 36.00 | 0.966 | 35.86 | 0.963 | 36.69 | 0.965 | 38.98 | 0.969 | **39.48** | **0.970** |
| 2 | 23.65 | 0.746 | 22.19 | 0.636 | 31.28 | 0.945 | 32.01 | 0.952 | 25.60 | 0.829 | 23.71 | 0.621 | 33.58 | 0.975 | 34.25 | 0.963 | 34.00 | 0.957 | 35.35 | 0.961 | 37.92 | 0.973 | **38.35** | **0.975** |
| 3 | 25.12 | 0.864 | 24.03 | 0.786 | 32.54 | 0.939 | 33.26 | 0.948 | 27.29 | 0.882 | 26.39 | 0.791 | 33.05 | 0.959 | 31.53 | 0.948 | 34.04 | 0.944 | 34.61 | 0.946 | 37.81 | 0.963 | **38.75** | **0.969** |
| 4 | 34.37 | 0.878 | 28.44 | 0.798 | 37.37 | 0.954 | 38.03 | 0.958 | 37.19 | 0.921 | 31.30 | 0.816 | 31.35 | 0.942 | 39.36 | 0.968 | 40.58 | 0.966 | 41.08 | 0.964 | 36.97 | 0.924 | **37.77** | **0.939** |
| 5 | 20.33 | 0.790 | 24.66 | 0.849 | 30.79 | 0.930 | 31.96 | 0.950 | 20.90 | 0.835 | 23.21 | 0.842 | 39.00 | 0.970 | 34.43 | 0.967 | 34.36 | 0.967 | 35.36 | 0.969 | 42.74 | 0.964 | **43.68** | **0.971** |
| 6 | 22.27 | 0.791 | 25.52 | 0.782 | 34.12 | 0.943 | 34.83 | 0.952 | 22.82 | 0.821 | 24.85 | 0.769 | 33.08 | 0.959 | 37.39 | 0.969 | 37.41 | 0.973 | 38.18 | 0.970 | 37.42 | 0.964 | **38.24** | **0.970** |
| 7 | 18.99 | 0.784 | 14.50 | 0.670 | 26.62 | 0.920 | 27.51 | 0.931 | 18.03 | 0.780 | 15.62 | 0.695 | 36.65 | 0.968 | 28.43 | 0.900 | 28.74 | 0.905 | 29.72 | 0.907 | 39.77 | 0.974 | **40.00** | **0.975** |
| 8 | 21.24 | 0.750 | 23.52 | 0.732 | 29.22 | 0.926 | 28.91 | 0.924 | 22.58 | 0.813 | 23.55 | 0.725 | 27.69 | 0.891 | 33.82 | 0.959 | 32.61 | 0.966 | 33.98 | 0.967 | 34.37 | 0.934 | **35.47** | **0.945** |
| 9 | 26.89 | 0.845 | 23.04 | 0.750 | 32.21 | 0.943 | 32.19 | 0.951 | 29.23 | 0.898 | 24.99 | 0.738 | 32.30 | 0.958 | 32.38 | 0.952 | 33.19 | 0.950 | 33.52 | 0.951 | 37.72 | 0.976 | **38.38** | **0.978** |
| 10 | 21.40 | 0.727 | 23.43 | 0.752 | 30.08 | 0.950 | 31.89 | 0.965 | 21.95 | 0.790 | 22.86 | 0.793 | 32.02 | 0.955 | 35.17 | 0.974 | 33.52 | 0.972 | 34.70 | 0.970 | 38.31 | 0.955 | **38.96** | **0.962** |
| Avg | 23.81 | 0.792 | 23.67 | 0.756 | 31.68 | 0.939 | 32.40 | 0.948 | 25.14 | 0.839 | 24.33 | 0.760 | 33.35 | 0.954 | 34.27 | 0.957 | 34.43 | 0.956 | 35.32 | 0.957 | 38.20 | 0.960 | **38.91** | **0.965** |

Table 1.The PSNR in dB and SSIM results of the test methods on 10 scenes

We use a machine with an Intel Xeon Gold 5218CPU and a NVIDIA RTX 3090 GPU to train our network. During the training process, the batch size is set to 5 and the learning rate is set to $10^{-4}$.

In the assessing process, peak signal-to-noise ratio (PSNR) and structural similarity (SSIM) are used to evaluate the reconstruction quality of the hyperspectral images. We assess spectral accuracy through comparing the spectral response curve with the ground truth.

### 4.2. Experiment on Synthetic Data

To assess the accuracy of our proposed method, we compare it with several mainstream methods in both the previous RCASSI system and our proposed SE-RCASSI system. In our simulation, we evaluate four algorithms: two optimization-based algorithms (TwIST and ADMM) and two deep learning-based algorithms (Unet-3D and TSA-Net), using a test set consisting of ten scenes from the KAIST dataset. Table 1 displays individual simulation results for each scene and the average results across all scenes. Particularly, SEnet-U represents the method that SEnet framework utilize Unet-3D as a module and reconstruct HSI with both Unet-3D module output and grayscale image. SEnet-UH refers to the method applying hybrid prior strategy to the SEnet-U.

The results reveal the superior performance of SEnet-UH, which achieves the highest PSNR with more than 1 dB improvement over other algorithms. Notably, the PSNR has been significantly increased when applied in SE-RCASSI system, which indicates the superiority of the system design and grayscale image fusion. Furthermore, Unet-3D and TSA-Net show significant improvements in reconstruction accuracy when deployed in SEnet framework, resulting in a 2 dB increase in PSNR compared to deploying original algorithm in SE-RCASSI system. These indicate that SEnet is able to exploit deeper prior information of the grayscale image than conventional algorithms. Additionally, the employment of narrow-scope prior learning greatly enhances the accuracy of reconstruction. The SEnet-UH achieves a 3 dB PSNR improvement and SEnet-TH achieves a 2 dB PSNR improvement than that of SEnet-T. These indicate that the narrow-scope prior greatly contributes to real world HSI reconstruction. In conclusion, these results emphasize the superiority of SEnet combined with hybrid prior strategy.

### 4.3. Hardware Prototype Implementation

We fabricate the prototype of SE-RCASSI. The internal structure of the system is shown in Fig.3 (a) and the final outlook of the spectral imager is shown in Fig.3(b). To verify the feasibility of the compact snapshot hyperspectral imager, we carried out the outdoor experiment as shown in Fig.3(c). The results of outdoor experiment are described in 4.5.

### 4.4. Experiment on Indoor scenes

Fig. 4 shows the reference RGB image, SE-RCASSI system measurements and reconstructed results of two scenes. We select four positions (a, b, c and d) in this scene to plot the spectral reconstruction curves. The spectral results of six algorithms (TwIST, TwIST with dual cameras, Unet-3D, Unet-3D with dual cameras, SEnet-U, SEnet-UH) are compared, as shown in Fig. 4. In the lower part of Fig. 4, we show 4 out of 27 spectral channels.

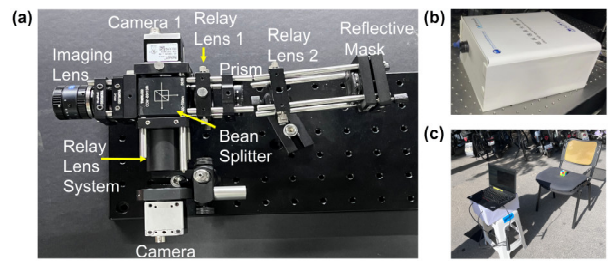

Fig 3. Hardware implementation of system. (a)The system proto type. (b) Outlook of capsulation. (c)Implementation of outdoor experiment.

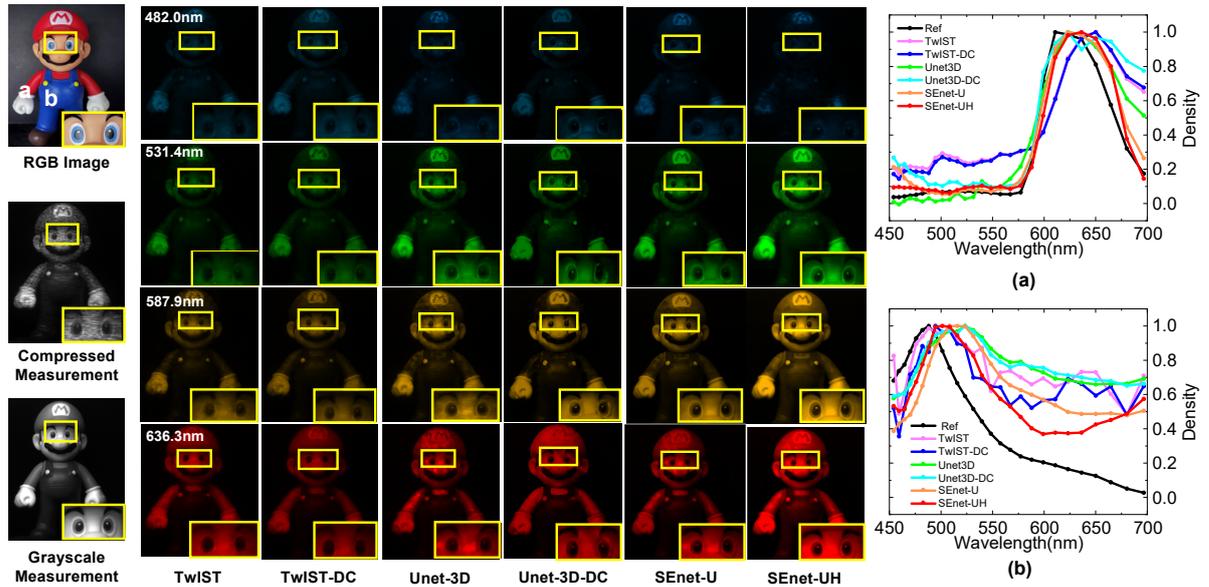

Fig 4. Real HSI scenes of Mario. Reconstructed results with Unet-3D, Unet-3D-DC (dual-camera), SEnet-U (SEnet connected with Unet-3D), SEnet-UH (Applying HPS to SEnet-U), TwIST and TwIST-DC (dual-camera). Two points marked by 'a' and 'b' in the RGB reference images are selected to plot the spectral curves.

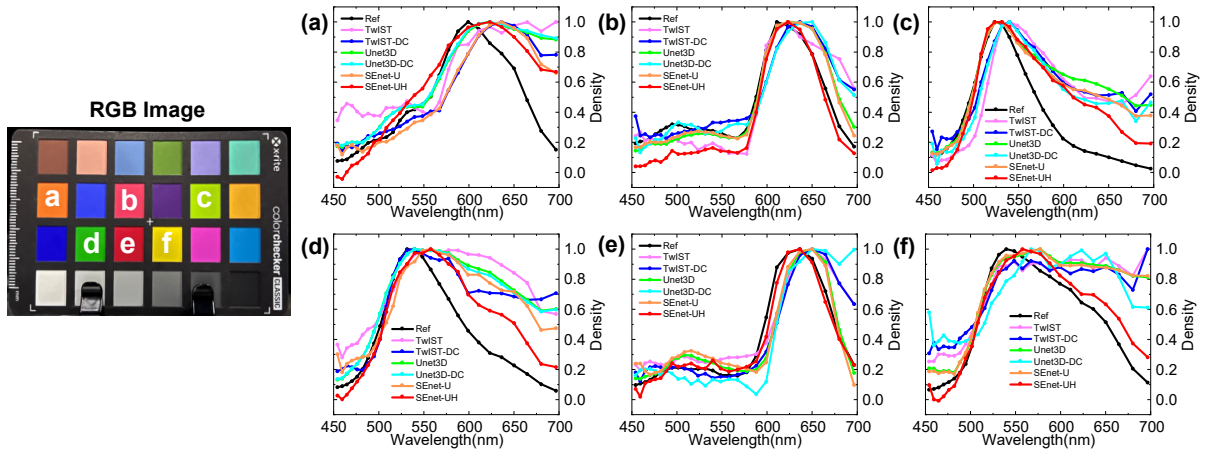

Fig 5. Real HSI scenes of color checker. Six points marked by 'a', 'b', 'c', 'd', 'e' and 'f' in the RGB reference images are selected to plot the spectral curves.

**Spatial results.** As shown in fig.4, the spatial results of TwIST and TwIST-DC suffer from a lot of blur and noise. Compared with TwIST and TwIST-DC, Unet-3D, Unet3D-DC, SEnet-U, SEnet-UH maintain more spatial details of the scene. For example, in the results of the latter four algorithms, the letter M on Mario's hat and its eyes are clearly reconstructed while the first two algorithms fail to reconstruct. In the comparison of the reconstruction results of the last four algorithms, SEnet-UH preserves more complete spatial details and has the least artifacts. The SEnet framework and the hybrid prior strategy help to effectively restore the spatial information in hyperspectral images.

**Spectral results.** As shown in Fig. 4 and Fig. 5, the spectral response curves of Unet-3D and SEnet-U are closer to the ground truth than that of TwIST and TwIST-DC. It indicates that Unet-3D and SEnet-U can restore the more accurate spectral information compared with the TwIST and TwIST-DC. Furthermore, SEnet-UH shows a significant improvement over the results of Unet-3D. The spectral response curve of SEnet-UH is closer to the ground truth than that of Unet-3D and SEnet-U.

accurate spectral reconstruction of all four methods, which indicates SEnet-UH's capability of outdoor application.

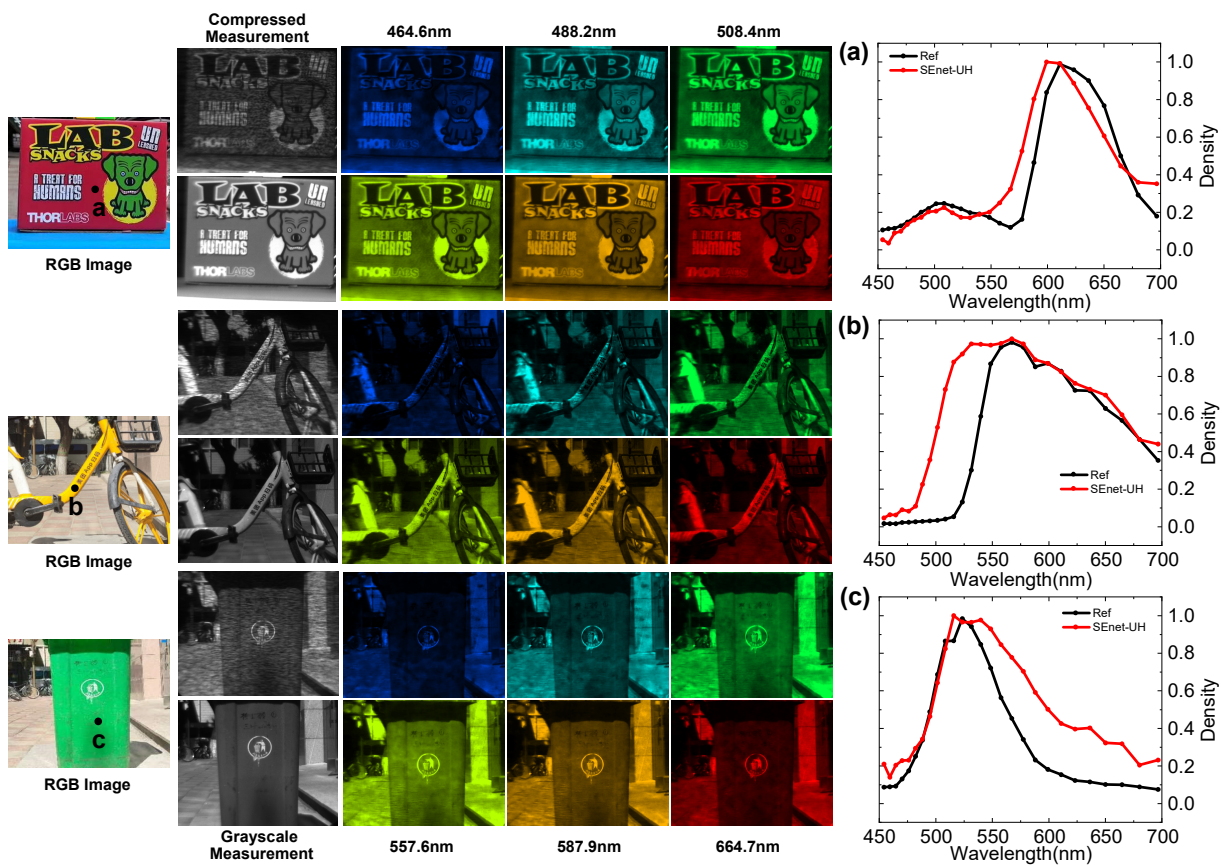

Fig 6. Real HSI of four outdoor scenes. Reconstructed results with SEnet-UH are shown. Three points marked by 'a', 'b' and 'c' in the RGB reference images are selected to plot the spectral curves.

4.5. Experiment on Outdoor Scenes

The former indoor experiments indicate the superior performance of SEnet-UH compared to other algorithms. In this part, we conduct the outdoor experiments to verify the feasibility of SE-RCASSI and SEnet when applied outdoor. Fig. 6 shows the reference RGB image, SE-RCASSI system measurements and reconstructed results of three outdoor scenes. We select three positions (a, b and c) in this scene to plot the spectral reconstruction curves. The spectral results of SEnet-UH are shown in Fig. 6. All scenes are captured by the system under direct sunlight.

As shown, SEnet-UH is able to reconstruct visual-pleasing hyperspectral images and preserve the spatial details of outdoor scenes. The spectral response results are also very close to the ground truth, which proves the SE-RCASSI system's feasibility in outdoor environments. Furthermore, SEnet-UH achieves the most

## 5. Conclusion

In conclusion, we propose SE-RCASSI system with high accuracy and high light efficiency. In addition, we propose the SEnet and hybrid prior strategy to fit this system. The SEnet with HPS achieves better reconstruction performance than previous algorithms. Notably, different HSI reconstruction algorithms can be deployed as the pre-reconstruction module in SEnet to attain better results. Furthermore, we have conducted numerical simulations and experiments to verify the feasibility of our method. In the simulation, SEnet-UH achieves more than 4dB improvement in PSNR compared with previous algorithms. In the indoor and outdoor experiments, the reconstruction performance by SEnet have shown outstanding quality of preserving spatial details and obtaining high spectral fidelity. In the future, we would do more efforts on the system generalization and network improvement.